\documentclass[12pt]{article}
\usepackage{amsfonts}
\usepackage{amsmath}
\usepackage{cite}

\newcommand{\eq}{\begin{equation}}
\newcommand{\feq}{\end{equation}}
\newcommand{\eqn}{\begin{eqnarray}}
\newcommand{\feqn}{\end{eqnarray}}
\newcommand{\arr}{\begin{eqnarray*}}
\newcommand{\farr}{\end{eqnarray*}}

\newcommand{\sw}{\stackrel{\star}{\wedge}}
\newcommand{\st}{\stackrel{\star}{\otimes}}

\newcommand{\At}{{\tilde A}}
\newcommand{\bt}{{\tilde b}}
\newcommand{\gt}{{\tilde g}}

\newcommand{\lp}{\left(}
\newcommand{\rp}{\right)}

\font\mybb=msbm10 at 12pt
\def\bb#1{\hbox{\mybb#1}}

\def\bR {\bb{R}}

\begin{document}

\begin{titlepage}
\begin{flushright}
IFUM-711-FT\\
hep-th/0204152
\end{flushright}
\vspace{.3cm}
\begin{center}
\renewcommand{\thefootnote}{\fnsymbol{footnote}}
{\Large \bf Noncommutative AdS Supergravity in three Dimensions}
\vskip 15mm
{\large \bf {S.~Cacciatori$^{1,3}$\footnote{cacciatori@mi.infn.it} and
L.~Martucci$^{2,3}$\footnote{luca.martucci@mi.infn.it}}}\\
\renewcommand{\thefootnote}{\arabic{footnote}}
\setcounter{footnote}{0}
\vskip 10mm
{\small
$^1$ Dipartimento di Matematica dell'Universit\`a di Milano,\\
Via Saldini 50, I-20133 Milano.\\

\vspace*{0.5cm}

$^2$ Dipartimento di Fisica dell'Universit\`a di Milano,\\
Via Celoria 16, I-20133 Milano.\\

\vspace*{0.5cm}

$^3$ INFN, Sezione di Milano,\\
Via Celoria 16,
I-20133 Milano.\\
}
\end{center}
\vspace{2cm}
\begin{center}
{\bf Abstract}
\end{center}
{\small 
In this paper we propose a generalization of ${\cal N}=4$
three dimensional AdS supergravity to the noncommutative case. This is a 
supersymmetric version of the results presented in 
\cite{Cacciatori:2002gq}. We show that it continues to admit an ${\cal
N}=4$ supersymmetric
solution which is the noncommutative couterpart of $AdS_3$ space. Some 
other solutions are also discussed.

}

\end{titlepage}

\section{Introduction}
As already known, open string theory in presence of a constant
background Neveu-Schwarz
$B-$ field gives rise, in the field theory limit, to noncommutative
gauge theories whose field's algebra is described by the Moyal product
\cite{Seiberg:1999vs}. Even if the particularly good properties of the 
Moyal product allows to treat these theories as ordinary field
theories on a certain smooth  manifold, noncommutativity changes drastically
the geometrical structure of this base manifold and geometrical objects
like the metric become difficult to define.   
Morever, the noncommutativity parameters 
$\theta^{\mu\nu}$ are
determined by $B_{\mu\nu}$ which generally behaves as a dynamical field;
this leads one to suspect 
that also noncommutativity (nonlocality) could become
dynamical. On the other hand, one can imagine a $D-$brane in a 
background $B-$ field 
with strings 
attached on it. These strings could close and escape from the brane giving
rise
to gravitational interactions. It should then be interesting to
understand how
field thories couple to gravity, and moreover the problem of a
consistent construction of gravity on noncommutative spaces becomes relevant
itself.
In particular, string theories being supersymmetric,
the construction of noncommutative supergravity must be also considered.\\
Till now, much work has been done to analyze
the structure of a noncommutative spacetime \cite{connesealtri} from a
mathematical point of view. It is plausible that if noncommutative gravity
were consistently built without reference to its
stringy origin, a better understanding of the space-time 
structure from a physical point of view, could be achieved.

In this paper we define a possible model of noncommutative
supersymmetric gravity extending the approach tried in \cite{Cacciatori:2002gq}
to the supersymmetric case. 
As in \cite{Cacciatori:2002gq} all the fields in our model happen to be real 
and a metric can be naturally defined. We also find the
supersymmetric extension of the invariances of
the action which reduce to diffeomorphisms in the
commutative limit.

Finally we analyse some solutions like the noncommutative 
analogue of $AdS_3$ space, and the BTZ black hole with 
zero mass and zero angular momentum. 
In particular, the noncommutative $AdS_3$ solution results to be a maximally 
supersymmetric solution, just like the commutative one.

\section{The action and its symmetries}

It is well known that three dimensional AdS supergravity, 
just like the non supersymmetric one, can be written  as a 
Chern-Simons theory \cite{Achucarro:vz}. Since Chern-Simons theory is 
well formulated also in the noncommutative case, we can start from it to 
produce a noncommutative version of three dimensional supergravity 
and, for that pourpose, the manifestly supersymmetric formulation employing
superalgebras and supertraces is particulary well suited.
As in the nonsupersymmetric case, use of the star product
forces the extension from the usual 
$su(1,1)\oplus su(1,1)$ algebra to $u(1,1)\oplus u(1,1)$ 
\cite{Cacciatori:2002gq}. Introducing two new
$u(1)$ gauge fields, the simplest supersymmetric extension
is now given by the superalgebra $u(1,1|1)\oplus u(1,1|1)$.
The supergravity action thus becomes\footnote{The conventions are given 
in appendix \ref{appendixA}}:

\eqn \label{action1}
S= \kappa \int \mathrm{Str} \left( \Gamma \sw d\Gamma +\frac{2}{3}\,
\Gamma\sw 
\Gamma\sw \Gamma\right)-
\kappa \int \mathrm{Str} \left( \tilde\Gamma \sw d\tilde\Gamma
+\frac{2}{3}\, 
\tilde\Gamma\sw 
\tilde\Gamma\sw \tilde\Gamma\right)\ ,
\feqn

where the 1-forms $\Gamma$ and $\tilde\Gamma$ are  $u(1,1|1)$ 
super connections and can be written in the following way (cf. appendix 
\ref{appendixA}):

\eqn \label{change1}
\Gamma=
\left( 
\begin{matrix}
A & \psi \cr 
-i\bar{\psi} & ig
\end{matrix}\right)\ ,\ 
\Gamma=
\left(
\begin{matrix}
\tilde A & \tilde\psi \cr 
-i\bar{\tilde\psi} & i\tilde g
\end{matrix}\right)\ ;
\feqn

here $A=A^A\tau_A$ and $\tilde{A} =\tilde{A}^A\tau_A$ are the bosonic
$u(1,1)$ 
1-form gauge fields, $g$ and $\tilde g$ are the $u(1)$
1-forms gauge fieldsassociated to the R-symmetry, whereas $\psi$ and 
$\tilde\psi$ 
are  complex 
spinorial 1-forms. In Eq.(\ref{action1}) $\kappa=-1/(16\pi G)$, where $G$
is the 
three dimensional Newton constant.
In terms of these fields, the action can be written in the following way:

\eqn \label{action2}
S&=& \kappa \int {\mathrm{Tr}} \lp A \sw dA +\frac 23 A \sw A \sw A \rp
-\kappa \int {\mathrm{Tr}} \lp \At \sw d\At +\frac 23 \At \sw \At \sw
\At\rp-
\cr &&-\kappa \int {\mathrm{Tr}} \lp g \sw dg +\frac{2i}{3} g \sw g \sw g
\rp
-\kappa \int {\mathrm{Tr}} \lp \gt \sw d\gt +\frac{2i}{3} 
\gt \sw \gt \sw \gt\rp+\cr
&&+2i\kappa\int \bar{\psi}\sw D\psi-
 2i\kappa\int \bar{\tilde\psi}\sw D\tilde\psi\ ,
\feqn

where $D\psi:= d\psi+A\sw \psi+i\psi\sw g$ and 
 $D\tilde\psi:= d\tilde\psi+\tilde A\sw \tilde\psi+i\tilde\psi\sw \tilde
g$. 
We have thus obtained an action which is the sum of the action resulting
in \cite{Cacciatori:2002gq} plus two Chern-Simons actions for the
R-symmetry
gauge fields $g$ and $\tilde g$, and two minimally coupled terms for the
gravitino fields $\psi$ and $\tilde\psi$.     
The action (\ref{action1}) is clearly invariant under
the infinitesimal $u(1,1|1)\oplus u(1,1|1)$ gauge trasformations

\eqn \label{gauge1}
&&\delta_\Lambda \Gamma = d\Lambda +\Gamma\star \Lambda-
\Lambda\star\Gamma\ ,\cr
&&\tilde\delta_{\tilde\Lambda}\Gamma = d\tilde\Lambda +
\tilde\Gamma\star\tilde\Lambda-
\tilde\Lambda\star\tilde\Gamma\ ,\cr
&&\tilde\delta_{\tilde\Lambda} \Gamma = \delta_\Lambda\tilde\Gamma=0\ .
\feqn

The infinitesimal gauge parameters $\lambda$ and $\tilde\lambda$ 
can be written in the 
explicit form

\eqn \label{change2}
\Lambda=
\left(
\begin{matrix}
\lambda^A\tau_A & \epsilon \cr 
-i\bar\epsilon & i\alpha
\end{matrix}\right)\ ,\qquad 
\tilde\Lambda=
\left(
\begin{matrix}
\tilde\lambda^A\tau_A & \tilde\epsilon \cr 
-i\bar{\tilde\epsilon} & i\tilde\alpha
\end{matrix}\right)\ ,
\feqn

so that the gauge transformations become

\eqn \label{gauge2}
\delta_\Lambda A &=& d\lambda+A\star\lambda-\lambda\star A
+i(\epsilon\st \bar\psi-\psi\st\bar\epsilon)\ ,\cr
\delta_\Lambda\psi &=& d\epsilon+
A\star\epsilon-\lambda\star \psi
+i\psi\star\alpha-i\epsilon\star g\ ,\cr
\delta_\Lambda g &=& d\alpha-\bar\psi\star\epsilon
+\bar\epsilon\star\psi+ig\star\alpha-i\alpha\star g\ ,
\feqn

and $\tilde A$, $\tilde\psi$ and $\tilde g$ transform exactly in the same
way 
under
$\tilde\delta_{\tilde\Lambda}$. The commutation rules for these 
transformations
are given in appendix \ref{appendixB}. \\
Following the same argument and notation employed in
Refs.\cite{Cacciatori:2002gq,cckmsz:2002}, it is
straightforward to show that the action (\ref{action1}) is also invariant
under the transformation
\eqn
\Delta_v^{\star}\Gamma := \delta_{i_v^{\star} \Gamma} \Gamma + 
i_v^{\star} D\Gamma\ ,\cr
\Delta_v^{\star}\tilde\Gamma := \tilde\delta_{i_v^{\star} \tilde\Gamma} 
\tilde\Gamma + 
i_v^{\star} D\tilde\Gamma\ ,\cr
\label{ncdiff}
\feqn
where  $D\Gamma = d\Gamma + \Gamma \sw \Gamma$. These transformations 
reduce to the usual diffeomorfisms ${\cal L}_v \Gamma$ in the 
$\theta^{\mu\nu}\rightarrow 0$ limit. This point has been discussed 
more deeply in the cited references.

Now we can make the same substitution as in
Ref.\cite{Cacciatori:2002gq},

\eqn \label{change3}
A &=& \omega +\frac el +\frac i2 b\,, \\
\At &=& \omega -\frac el +\frac i2 \tilde b\,,
\feqn

with

\eqn
\omega = \omega^a \tau_a \ , \qquad e = e^a \tau_a \ , \qquad b=b
\mathbb{I}
\feqn

so that the supergravity action can be written in the more familiar
way 

\eqn
S &=& -\kappa \int \epsilon_{abc} \lp e^a \sw R^{bc} +\frac{1}{3l^2}
      e^a \sw e^b \sw e^c \rp \nonumber \\
&& - \frac{l\kappa}{2} \int \lp b\sw db +\frac i3 b\sw b\sw b \rp +
     \frac{l\kappa}{2} \int \lp \bt\sw d\bt +\frac i3 \bt\sw \bt\sw \bt
\rp
     \nonumber \\
&& + \frac{i\kappa}{2} \int \eta_{ab} \lp e^a \sw \omega^b +\omega^a \sw
e^b
     \rp \sw \lp b +\bt \rp \nonumber \\
&& + \frac{il\kappa}{2} \int \eta_{ab} \lp \omega^a \sw \omega^b
+\frac{1}{l^2}
     e^a \sw e^b \rp \sw \lp b -\bt \rp+\cr
&& +  \kappa \int {\mathrm{Tr}} \lp g \sw dg +\frac{2i}{3} g \sw g \sw g
\rp
-\kappa \int {\mathrm{Tr}} \lp \gt \sw d\gt +\frac{2i}{3} 
\gt \sw \gt \sw \gt\rp+\cr
&&+2i\kappa\int \bar{\psi}\sw D\psi-
 2i\kappa\int \bar{\tilde\psi}\sw D\tilde\psi\ ,   
\feqn

where we have introduced

\eqn
&& R^{ab} = d\omega^{ab} +\frac 12 \lp \omega^a_{\,\,\, c} \sw \omega^{cb}
         - \omega^b_{\,\,\, c} \sw \omega^{ca} \rp\ ,\cr
&& T^a = de^a +\frac 12 \lp \omega^a_{\,\,\, b} \sw e^b +e^b \sw
      \omega_b^{\,\,\, a} \rp\ ,
\feqn

with $\omega^{ab}=\epsilon^{abc}\omega_c$.

The natural definition of a  metric is

\eqn 
G_{\mu\nu}=e^a_\mu \star e^b_\nu \eta_{ab}=g_{\mu\nu  }+ib_{\mu\nu}\ ,
\feqn

where

\eqn
g_{\mu\nu}=\frac12 \eta_{ab}\{ e^a_\mu,e^b_\nu\}_\star
\feqn

is real and symmetric, and reduces to the usual expression of the metric in 
the commutative case, whereas

\eqn
b_{\mu\nu}=-\frac{i}{2}\,\eta_{ab}[e^a_\mu,e^b_\nu]_\star
\feqn

is real and antisymmetric, and vanishes for $\theta^{\mu\nu}=0$.

\section{The BPS solutions}

In this section we will look for supersymmetric solutions of
our
model, obtaining the noncommutative analogues of some classical
supersymmetric
solutions of commutative $AdS_3$ supergravity, namely the ``fuzzy'' $AdS_3$ and
the massless nonrotating BTZ
black hole. Finally, we will find a
generalization
of the massless BTZ black hole, including 
$U(1)$ gauge fields $b$ and $g$. To this end, we will follow the 
noncommutative version of the standard BPS construction.

As in the commutative case, one puts the fermionic fields to zero 
and looks for pure bosonic
solutions of the equations of motion, satisfing the conditions 
$\delta_\epsilon\psi=\delta_{\tilde\epsilon}\tilde\psi=0$ for some
$\epsilon$,
to end with a residual supersymmetry.
At the moment, we can restrict to the untilded sector, the tilded one
behaving
identically. From the second line of Eq.(\ref{gauge2}), our conditions read

\eqn \label{killing}
d\epsilon+A\star\epsilon-i\epsilon\star g=0\ .
\feqn

From $d^2=0$ one gets the consistence condition $(dA+A\sw A)\star \epsilon
=i\epsilon\star(dg+ig\sw g)$, which is always satisfied on-shell 
(when $\psi=0$). So, our noncommutative Killing equation
Eq.(\ref{killing}) 
seems to be always solvable, at least locally, when the fields 
$A$ and $g$ are on-shell. Note also from Eq.(\ref{action2})
that  when $\psi=0$, 
the fields $A$ and $g$ are completely 
decoupled, resulting in an $AdS_3$ noncommutative gravity 
(see \cite{Cacciatori:2002gq}) plus two $U(1)$
Chern-Simons theories. However from the Killing equation
Eq.(\ref{killing}) one 
sees that supersymmetry conditions restore a dependence between the $g$
and
the $A$ fields.

In order to express the Killing conditions (\ref{killing}) in terms of the 
gravitational fields, as defined in Eq.(\ref{change3}), it is convenient
to 
introduce the notation $\hat\delta_{(\Lambda_1,\Lambda_2)}:=
\delta_{\Lambda_1+\Lambda_2}+\tilde\delta_{\Lambda_1-\Lambda_2}$, and to put
$\psi_1=\psi+\tilde\psi$,
$\psi_2=\psi-\tilde\psi$, $\epsilon=\epsilon_1+\epsilon_2$ and
$\tilde\epsilon=\epsilon_1-\epsilon_2$.
Now the generic two-parameters supersymmetric
transformations of the fermions can be written as

\eqn \label{susy2}
\hat\delta_{(\epsilon_1,\epsilon_2)}\psi_1 &=& 2d\epsilon_1 +
\omega^a\gamma_a\star\epsilon_1+l^{-1}e^a\gamma_a\star\epsilon_2 +
\frac i2 (b+\tilde b)\star\epsilon_1+
\frac i2 (b-\tilde b)\star\epsilon_2 \cr &-& 
i\epsilon_1\star(g+\tilde g)\-
i\epsilon_2\star(g-\tilde g)\ ,\cr 
\hat\delta_{(\epsilon_1,-\epsilon_2)}\psi_2 &=& 2d\epsilon_2 +
\omega^a\gamma_a\star\epsilon_2+
l^{-1}e^a\gamma_a\star\epsilon_1+
\frac i2 (b+\tilde b)\star\epsilon_2+
\frac i2 (b-\tilde b)\star\epsilon_1 \cr &-&
i\epsilon_2\star(g+\tilde g)\-
i\epsilon_1\star(g-\tilde g)\ .
\feqn

We now show that there is a maximally supersymmetric solution,
which is the noncommutative correspondent of the $AdS_3$ space and that
for 
brevity we will call  the ``fuzzy'' $AdS_3$\footnote{It would be 
interesting to see if there really exists fuzzy construction to which 
this solution corresponds.}. To this end let us
fix, once for all, the coordinates $(t,r,\phi)$ as parametrizing
$\bR \times \mbox{\bf{C}}$,
where $\mbox{\bf{C}}$ is a "fuzzy cylinder", whose spatial
coordinates $(r,\phi)$
satisfy the non commutativity condition
\eqn  \label{parentesi}
\left[ r,\phi \right] =i\theta \ .
\feqn
It is then 
easy to see that the ansatz
\eqn
e^0 &=&\sqrt{\frac{r^2}{l^2}+1} \,dt=:N(r) dt \cr
e^1 &=& N(r)^{-1} dr  \\
e^2 &=& r d\phi \nonumber
\feqn
and
\eqn
\omega^0 &=& -N(r) \,d\phi \cr
\omega^1 &=& 0  \\
\omega^2 &=& \frac r{l^2} \,dt\ , \nonumber
\feqn
is a solution of the equations of 
motion when all the other fields are set to zero. 
This solution corresponds to a diagonal symmetric real metric $G_{\mu\nu}$ 
so that the corresponding imaginary part is $b_{\mu\nu}=0$ whereas
$g_{\mu\nu}$ is the same as
in the commutative case
\eqn
ds^2=-\left( \frac{r^2}{l^2}+1\right)dt^2+
\left( \frac{r^2}{l^2}+1\right)^{-1}dr^2+r^2d\phi^2
\feqn

To solve eq. (\ref{susy2}), it is convenient to consider $\epsilon_1$
proportional to $\epsilon_2$ and so, without loss of generality, treat
separately the two
independent choices
$\epsilon_1 =\alpha~~\epsilon_2 = \epsilon$, with $\alpha=\pm
1$;
this is equivalent to put $\tilde \epsilon$ or $\epsilon$ equal to zero 
respectively (and can be understood in terms of  the two possible 
representations of the
Dirac matrices in three dimensions \cite{Coussaert:1993}).

The BPS equations are then
\eqn
2\partial_t \epsilon -\frac r{l^2} \gamma_2 \star \epsilon +\frac
{\alpha}l
N(r) \gamma_0 \star \epsilon &=& 0 \cr
2\partial_r \epsilon +\frac {\alpha}l N(r)^{-1} \gamma_1 \star \epsilon
&=& 0  \\
2\partial_{\phi} \epsilon -N(r) \gamma_0 \star \epsilon
+\alpha \frac rl \gamma_2 \star \epsilon &=& 0\ . \nonumber
\feqn
By means of the relations $\gamma_0 \gamma_1 \gamma_2 =1$ and
$(N(r)+1)^{\frac 12}(N(r)-1)^{\frac 12} =\frac rl$, it can be seen that the
Killing
spinors are
\eqn \label{susyads3}
\epsilon = \left[ (N(r)+1)^{\frac 12} -\alpha (N(r)-1)^{\frac 12} \gamma^1
\right]
\star \left[ \cos \left( \frac {\phi}{2} -\frac{\alpha t}{2l}\right)
-\sin \left( \frac {\phi}{2} -\frac{\alpha t}{2l}\right) \gamma^0 \right]
\xi
\feqn
for every choice of $\alpha$ and constant spinor $\xi$.
As in the commutative case \cite{Coussaert:1993} there are four
independent 
generators of supersymmetries. \par

Next we generalize to the case of a static BPS black hole. Let us make the  
ansatz

\eqn
e^0 =e^{A(r)} dt \ , \ \ \ e^1 =e^{B(r)} dr \, \ \ \ e^2 =e^{C(r)}
d\phi \ .
\feqn
We consider again the case in which $b_{\mu\nu}=0$, and $g_{\mu\nu}$
is given by

\eqn
ds^2=-e^{2A(r)}dt^2+e^{2B(r)}dr^2+e^{2C(r)}d\phi^2\ .
\feqn 

Taking all the other fields but the spin connection, equal to zero
one finds
\eqn
\omega^0 =-e^{-B} C' e^2 \ , \ \ \ \omega^1 =0 \ , \ \ \
\omega^2 =-e^{-B} A' e^0 \ .
\feqn

The fuzzy $AdS_3$ found above clearly belongs to this more general 
class. In order to study the BPS equations (\ref{susy2}), 
we closely follow \cite{Coussaert:1993} and assume
that $\epsilon$ depends on $r$ only. In this way we are selecting 
particular coordinate systems so that for example the fuzzy $AdS_3$ case 
is excluded by this condition, as it is clear from Eq.(\ref{susyads3}).
It would be interesting to study the possible equivalence of different 
solutions under the transformations (\ref{ncdiff}). \par 
Under the previous assumption, the BPS equations (\ref{susy2})
read 
\eqn \label{BPS}
2 \partial_r \epsilon +\alpha \frac 1l e^1 \gamma_1 \epsilon &=& 0 \ , \cr
-e^{-B} A' \gamma_2 \epsilon +\alpha \frac 1l \gamma_0 \epsilon &=& 0 \ ,
\\
-e^{-B} C' \gamma_0 \epsilon +\alpha \frac 1l \gamma_2 \epsilon &=& 0 \ .
\nonumber
\feqn
The first equation of (\ref{BPS}) requires $\epsilon^{\pm} =Q(r) 
\epsilon^{(0)}_{\pm}$
where
$Q(r)$ is a scalar function of $r$ and $\epsilon_{\pm}^{(0)}$ is a constant
spinor
satisfying
\eqn
\gamma_1 \epsilon_{\pm}^{(0)} =\pm \epsilon_{\pm}^{(0)} \ .
\feqn
Multiplying the second and the third equations of (\ref{BPS}) by $\gamma_2$ and
$\gamma_0$ respectively and choosing $\alpha$ appropriately,
it is now easy to obtain as a complete solution\footnote{in the gauge
$e^C=r$.},
\eqn  \label{firstsol}
e^0 =\frac rl dt \ , \ \ \ e^1 =\frac lr dr \, \ \ \ e^2 =r d\phi \
,
\feqn
which corresponds to the metric

\eqn \label{trenta}
ds^2=-\frac{r^2}{l^2}dt^2+\frac{l^2}{r^2}dr^2+ r^2 d\phi^2
\feqn

and has  supersymmetry generators
\eqn
\epsilon^{\pm} =\sqrt{\frac rl}
\left( \begin{array}{c} 1 \\ \pm 1 \end{array} \right) \ .
\feqn
This is a fuzzy version of the BTZ black hole with zero mass and zero
angular
momentum \cite{Coussaert:1993}. Note that to obtain this solution one has
to
make peculiar choices of $\alpha$: requiring a positive sign for $e^{B(r)}$
needs $\alpha$ to have a sign which is opposite to the one of the 
given eigenvalue of $\gamma_1$ ; as 
before,
this is equivalent to choose only one of the two possible representations of
the
Dirac matrices in three dimensions \cite{Coussaert:1993}.\par
One can easily obtain a further solution, which is similar to (\ref{trenta}),
but has nonvanishing $U(1)$ gauge fields. Let us take
\eqn
b=\tilde b =2g =2\tilde g =b_0 [\phi dr +\beta r d\phi]
\feqn
with $b_0$ and $\beta$ constants. Then it is easy to see that all the
equations
 of
motion are satisfied if $\beta=1/(1+\theta b_0 )$ and that a new solution
is given by
\eqn  \label{secondsol}
e^0 =\frac rl dt \ , \ \ \ e^1 =\frac lr dr \, \ \ \ e^2 = r d\phi \ ,\cr
b=\tilde b = b_0 \phi dr+\frac {b_0}{1+\theta b_0} r d\phi \ , \\
g=\tilde g = \frac{b_0}2 \phi dr+\frac {b_0}{1+\theta b_0} \frac r2 d\phi
\ ,
\nonumber
\feqn
with supersymmetry generators:
\eqn
\epsilon^{\pm} =\left(\frac rl \right)^{\frac{1}{2+b_0 \theta}}
\left( \begin{array}{c} 1 \\ \pm 1 \end{array} \right) \ .
\feqn

In the commutative case this solution reduces to a zero mass BTZ black
hole plus four completely decoupled $U(1)$ gauge sectors 
given by
\eqn
b=\tilde b =2g =2\tilde g =b_0 [\phi dr + r d\phi]=
-ie^{-ib_0\phi r}d e^{ib_0\phi r}\ ;
\feqn
the latter can then be written at least locally,
as pure gauge solutions but not globally, and can thus
be considered as non trivial statical field configurations.
For example, 
the Wilson loop of $b$ along the closed curve 
$\gamma:\pi\mapsto (t_0,r_0,\phi)$
where $t_0$ and $r_0$ are constants and $\phi\in [0,2\pi]$,
is given by 
\eqn \label{Wilson}
W_\gamma[b]={\cal P}e^{i\oint_\gamma b}= exp(ib_0 2\pi r_0 ) \ ,
\feqn
so that the parameter $b_0$ can be 
considered as a coordinate in the moduli space of this kind of solutions.
In the noncommutative case these $U(1)$ gauge fields acquire
pure noncommutative couplings with the supergravity sector of the theory
which forces the fields $b,\tilde b$ and $g, \tilde g$ to be equal, whereas
in the commutative limit they are completely decoupled and need only satisfy
the zero curvature condition separately. Moreover noncommutativity
requires a modification of the 
dependence from the moduli coordinate $b_0$.
The fact that for these solutions the noncommutative curvature of each $U(1)$
is equal to zero allows to apply the prescription given by Alekseev and
Bytsko to calculate the noncommutative monodromies defined in 
\cite{Alekseev:2000td}, which reduce to ordinary Wilson loops in the case of
$U(1)$ commutative gauge theory\footnote{Here we have 
adapted the definition so as to take care
of the fact that in our case the gauge action is on the right, contrarily 
to \cite{Alekseev:2000td}}:
\eqn
M=G^\prime \star G^{-1}_\star\ , 
\feqn
where 
\eqn
G=Ae^{\frac{ib_0}{1+b_0\theta/2}r\phi} 
\feqn
satisfies the condition $dG=iG\star b$, $G^\prime(r,\phi):=G(r,\phi+2\pi)$
and $A$ is a constant that must be chosen in order to have 
$G^\dagger \star G=1$. For example, at second order in $\theta$ we have
\eqn
A=1+\frac{\theta^2 b_0^2}{2}+O(\theta^3)
\feqn
and so, after a straightforward calculation up to second order in $\theta$,
the noncommutative monodromy, 
associateted to the Wilson loop (\ref{Wilson}),  results in:
\eqn
M=e^{2\pi i b_0 r}\left[ 1+\theta^2\left( \frac i2 \pi b_0^3 r -\frac 12 \pi^2
b_0^4 r^2\right)\right]+O(\theta^3)\ .
\feqn 
Note that these field configurations have singularities for 
$b_0\theta^{-1}=-1,-2$: first of all this shows that there are solutions
of the commutative theory which have no noncommutative counterpart, but
as in the case of
Yang-Mills theories \cite{Seiberg:1999vs}, this also suggests that, 
in general, 
singular solutions
can arise, with a $\theta$ dependent singularity and that one should expect
these to have no analogue in the correspondent commutative theory.

\vspace{15 cm}


\bigskip 
{\bf Acknowledgements}
\small
We are grateful to D.~Klemm, D.~Zanon and G.~Berrino for useful discussions and
comments.
This work was partially supported by INFN, MURST and
by the European Commission RTN program
HPRN-CT-2000-00131, in which the authors are
associated to the University of Torino.

\newpage

\normalsize

\appendix

\section{Conventions} \label{appendixA}

An element $M$ of the Lie algebra u(1,1) is a complex $2\times2$ matrix
satisfing

\begin{equation}
M^i{}_j = - \eta_{jk} (M^k{}_l)^\ast \eta^{li}\,, \label{u11}
\end{equation}

where the $\ast$ denotes complex coniugation and
$\eta = {\mathrm{diag}}(-1,1)$. We choose as u(1,1) generators

\begin{eqnarray}
\tau_0 &=& \frac 12\left(\begin{array}{cc} i & 0 \\ 0 & -i \end{array}
           \right)\,, \qquad
\tau_1 = \frac 12\left(\begin{array}{cc} 0 & 1 \\ 1 & 0
\end{array}\right)\,,
         \nonumber \\ && \nonumber \\
\tau_2 &=& \frac 12\left(\begin{array}{cc} 0 & -i \\ i & 0 \end{array}
           \right)\,, \qquad
\tau_3 = \frac 12\left(\begin{array}{cc} i & 0 \\ 0 & i
\end{array}\right)\,.
         \label{generators}
\end{eqnarray}

They are normalized according to

\begin{equation}
{\mathrm{Tr}}(\tau_A\tau_B) = \frac 12 \eta_{AB}\,,
\end{equation}

where $(\eta_{AB}) = {\mathrm{diag}}(-1,1,1,-1)$ is the inner product
on the Lie algebra. The generators (\ref{generators}) satisfy the relation
(\ref{u11}). Further, if $a,b,c$ assume the values $0,1$ and $2$, then
the following relations hold:

\begin{eqnarray}
\left[ \tau_a ,\tau_b \right] &=& -\epsilon_{abc}\tau^c\,, \\
\left[ \tau_a ,\tau_3 \right] &=& 0\,, \\
\tau_a \tau_b &=& -\frac 12 \epsilon_{abc}\tau^c -\frac i2
\eta_{ab} \tau_3\,, \\
{\mathrm{Tr}}(\tau_a\tau_b\tau_c) &=& -\frac 14 \epsilon_{abc}\,, \\
{\mathrm{Tr}}(\tau_a\tau_b\tau_3) &=& \frac i4 \eta_{ab}\,,
\end{eqnarray}

where $(\eta_{ab}) = {\mathrm{diag}}(-1,1,1)$.
Furthermore we defined $\epsilon_{012} = 1$.

A general element of the superalgebra $\Omega\in u(1,1|1)$ is a $3\times
3$
supermatrix which must satisfy the condition

\begin{equation}
\Omega^a{}_b = - \eta_{bc}(\Omega^c{}_d)^\ast\eta^{da}\,.
\end{equation} 

and can be written as

\eqn
\Omega=
\left(
\begin{matrix}
M & \zeta \cr 
-i\bar{\zeta} & i\gamma
\end{matrix}\right)\ .
\feqn

where $M\in u(1,1)$ and $\gamma$ constitute the bosonic part of $\Omega$,
while $\zeta$   is a complex fermionic 2-dimensional vector in the
spinorial representation of $so(2,1)\simeq su(1,1)$ given by the
following representation of the three dimensional Clifford algebra 
$\{ \gamma_a ,\gamma_b\}=2\eta_{ab}$:

\eqn
\gamma_a:=2\tau_a & \Rightarrow& 
\gamma_0=\left( 
\begin{matrix} i & 0 \cr 0 & -i
\end{matrix}\right)\ ,\ \gamma_1=\left( 
\begin{matrix} 0 & 1 \cr 1 & 0
\end{matrix}\right)\ , \gamma_2=\left( 
\begin{matrix} 0 & -i \cr i & 0
\end{matrix}\right)\ ,
\feqn

which satisfies $\gamma^\dagger_a=\gamma_0\gamma_a\gamma_0,\ 
\gamma_a^\dagger \gamma_a=1$ and 
$[\gamma_a,\gamma_b]=-2\epsilon_{abc}\gamma^c$. We have also introduced
the 
notation $\bar \zeta=\zeta^\dagger\gamma_0$.

\section{The supersymmetry algebra}\label{appendixB}

We can restrict to analyze the transformation 
of $\Gamma$, the ``tilded'' sector being identical; the gauge
transformations 
(\ref{gauge1}) obey the commutatition rules $[\delta_{\Lambda_1},
\delta_{\Lambda_2}]=-\delta_{[\Lambda_1,\Lambda_2]}$. In terms of the
new fields introduced in Eq.(\ref{change1}),(\ref{change2}), 
they can be written
in the following way:

\eqn
[\delta_{\Lambda_1},\delta_{\Lambda_2}]A &=& -d[\lambda_1
,\lambda_2]_\star
+i\, d(\epsilon_1\st\bar\epsilon_2-\epsilon_2\st\bar\epsilon_1)+
[[\lambda_1 ,\lambda_2]_\star , A]_\star- \cr
&&- i[(\epsilon_1\st\bar\epsilon_2-
\epsilon_2\st\bar\epsilon_1), A]_\star -
i(\lambda_1\star \epsilon_2 - \lambda_2\star\epsilon_1)\st \bar\psi+\cr
&& + i\psi\st (\bar\epsilon_1\star\lambda_2 
-\bar\epsilon_2\star\lambda_1)+(\epsilon_1\star \alpha_2 
- \epsilon_2\star\alpha_1)\st \bar\psi-\cr
&&-\psi\st (\alpha_1\star\bar\epsilon_1 -\alpha_2\star\bar\epsilon_2)\
,\cr
&&\cr
[\delta_{\Lambda_1},\delta_{\Lambda_2}]\psi &=&
-d(\lambda_1\star\epsilon_2
-\lambda_2\star\epsilon_1)-i\, d(\epsilon_1\star\alpha_2
-\epsilon_2\star\alpha_1)+[\lambda_1,\lambda_2]_\star\star\psi-\cr
&&-i(\epsilon_1\st \bar\epsilon_2-\epsilon_2\st \bar\epsilon_1)\star\psi
+i(\lambda_1\star\epsilon_2-\lambda_2\star\epsilon_1)\star g -\cr 
&& -(\epsilon_1\star\alpha_2-\epsilon_2\star\alpha_1)\star g- 
A\star(\lambda_1\star\epsilon_2-\lambda_2\star \epsilon_1)-\cr
&&-i A\star(\epsilon_1\star\alpha_2-\epsilon_2\star \alpha_1)+
i\psi\star
(\bar\epsilon_1\star\epsilon_2-\bar\epsilon_2\star\epsilon_1)+\cr
&& +\psi\star (\alpha_1\star\alpha_2-\alpha_2\star\alpha_1)\ ,\cr
&&\cr
[\delta_{\Lambda_1},\delta_{\Lambda_2}]g &=& 
d (\bar\epsilon_1\star\epsilon_2-\bar\epsilon_2\star\epsilon_1)-
i\, d[\alpha_1,\alpha_2]_\star -(\bar\epsilon_1\star\lambda_2 
-\bar\epsilon_2\star\lambda_1)\star\psi-\cr
&&-i (\alpha_1\star\bar\epsilon_1 
-\alpha_2\star\bar\epsilon_2)\star\psi 
-i[(\bar\epsilon_1\star\epsilon_2-\bar\epsilon_2\star\epsilon_1),g]_\star-
[[\alpha_1,\alpha_2]_\star,g]_\star+\cr
&&+\bar\psi\star (\lambda_1\star\epsilon_2-\lambda_2\star\epsilon_1)+
i\bar\psi\star(\epsilon_1\star\alpha_2-\epsilon_2\star\alpha_1)\ .
\feqn


\newpage

\begin{thebibliography}{99}

\bibitem{Cacciatori:2002gq}
S.~Cacciatori, D.~Klemm, L.~Martucci and D.~Zanon,
``Noncommutative Einstein-AdS gravity in three dimensions,''
arXiv:hep-th/0201103.

\bibitem{connesealtri}
A.~Connes, ``Noncommutative Geometry'', Academic Press;
A.~Connes,
``Noncommutative Geometry And Reality,''
J.\ Math.\ Phys.\  {\bf 36} (1995) 6194; A.~Connes,
``Gravity coupled with matter and the foundation of 
non-commutative  geometry,''
Commun.\ Math.\ Phys.\  {\bf 182} (1996) 155
[arXiv:hep-th/9603053];
A.~H.~Chamseddine and A.~Connes,
``Universal Formula For Noncommutative Geometry Actions: Unification 
Of Gravity And The Standard Model,''
Phys.\ Rev.\ Lett.\  {\bf 77} (1996) 4868;
A.~H.~Chamseddine and A.~Connes,
``A universal action formula,''
Phys.\ Rev.\ Lett.\  {\bf 77} (1996) 4868
[arXiv:hep-th/9606056];
A.~H.~Chamseddine and A.~Connes,
``The spectral action principle,''
Commun.\ Math.\ Phys.\  {\bf 186} (1997) 731
[arXiv:hep-th/9606001]; G.~Landi and C.~Rovelli,
``Gravity from Dirac eigenvalues,''
Mod.\ Phys.\ Lett.\ A {\bf 13} (1998) 479
[arXiv:gr-qc/9708041]; G.~Landi and C.~Rovelli,
``General Relativity in terms of Dirac Eigenvalues,''
Phys.\ Rev.\ Lett.\  {\bf 78} (1997) 3051
[arXiv:gr-qc/9612034]. 

\bibitem{Seiberg:1999vs}
N.~Seiberg and E.~Witten,
JHEP {\bf 9909} (1999) 032
[arXiv:hep-th/9908142].

\bibitem{Minwalla:1999px}
S.~Minwalla, M.~Van Raamsdonk and N.~Seiberg,
JHEP {\bf 0002} (2000) 020
[arXiv:hep-th/9912072].

\bibitem{Chamseddine:2000si}
A.~H.~Chamseddine,
Phys.\ Lett.\ B {\bf 504} (2001) 33
[arXiv:hep-th/0009153].

\bibitem{Chamseddine:2000zu}
A.~H.~Chamseddine,
Commun.\ Math.\ Phys.\  {\bf 218} (2001) 283
[arXiv:hep-th/0005222].

\bibitem{Chamseddine:2000rj}
A.~H.~Chamseddine,
Int.\ J.\ Mod.\ Phys.\ A {\bf 16} (2001) 759
[arXiv:hep-th/0010268].

\bibitem{Damour:1992bt}
T.~Damour, S.~Deser and J.~G.~McCarthy,
Phys.\ Rev.\ D {\bf 47} (1993) 1541
[arXiv:gr-qc/9207003].

\bibitem{Moffat:fc}
J.~W.~Moffat,
J.\ Math.\ Phys.\  {\bf 36} (1995) 3722
[Erratum-ibid.\  {\bf 36} (1995) 7128];\\
Phys.\ Lett.\ B {\bf 491} (2000) 345
[arXiv:hep-th/0007181].

\bibitem{cckmsz:2002}
S.~Cacciatori, A.~H.~Chamseddine, D.~Klemm, L.~Martucci, W.~A.~Sabra and
D.~Zanon,
[hep-th/0203038].

\bibitem{Achucarro:vz}
A.~Achucarro and P.~K.~Townsend,
Phys.\ Lett.\ B {\bf 180} (1986) 89.

\bibitem{Witten:1988hc}
E.~Witten,
Nucl.\ Phys.\ B {\bf 311} (1988) 46.

\bibitem{Banados:2001xw}
M.~Banados, O.~Chandia, N.~Grandi, F.~A.~Schaposnik and G.~A.~Silva,
Phys.\ Rev.\ D {\bf 64} (2001) 084012
[arXiv:hep-th/0104264].

\bibitem{Coussaert:1993}
O.~Coussaert and M.~Henneaux,
Phys.\ Rev.\ Lett. {\bf 72} (1994) 183
[arXiv:hep-th/9310194].

\bibitem{Alekseev:2000td}
A.~Y.~Alekseev and A.~G.~Bytsko,
Phys.\ Lett.\ B {\bf 482} (2000) 271
[arXiv:hep-th/0002101].

\end{thebibliography}
\end{document}